\documentclass{ws-procs975x65}

\def\beq{\begin{equation}}
\def\eeq{\end{equation}}

\begin{document}

\title{Quantum fluctuations in FRLW space-time}
\author{Yevgeniya  Rabochaya$^*$ }

\address{Dipartimento di Fisica, Universit\`a di Trento, \\
Centro INFN-TIFPA, Trento\\
Via Sommarive 14, 38123 Povo, Italia\\
$^*$E-mail: yevgeniya.rabochay@unitn.it\\}

\begin{abstract}
In this paper we study a quantum field theoretical approach, where a quantum probe is used to investigate the properties of generic non-flat FRLW space time. The fluctuations related to a massless conformal coupled scalar field defined on a space-time with horizon is identified with a probe and the procedure to measure the local temperature is presented.

\end{abstract}


\bodymatter


\section{Introduction}

The Hawking radiation \cite{Haw} is one of the most robust and important predictions of
quantum field theory in curved space-time.
Here we would like to study some (local) properties of a generic Friedmann-Lemaitre-Robertson-Walker (FLRW) 
space-time with non-flat topology.

Let us remind some basic facts about the formalism. Any spherically symmetric four dimensional metric can be expressed in the form:   
\begin{equation}
ds^2 =\gamma_{ij}(x^i)dx^idx^j+ {\mathcal R}^2(x^i) d\Omega_2^2\,,\qquad i,j \in \{0,1\}\,,
\end{equation}
with $\gamma_{ij}(x^i)$ a tensor describing a two-dimensional space-time with coordinates $x^i$, $\mathcal R(x^i)$ being the ``areal radius'' and $d\Omega^2$ encoding the metric of a two-dimensional sphere orthogonal respect to the first one.

The dynamical trapping horizon -if exists- is located in the
correspondence of 
\begin{equation} 
\chi(x^i)\Big\vert_H = 0\,,\quad\partial_i\chi(x^i)|_H\gneq0\,, 
\quad
\chi(x^i)=\gamma^{ij}(x^i)\partial_i  {\mathcal R}(x^i)\partial_j  {\mathcal R}(x^i)\,. 
\end{equation} 
Thus, one may define the quasi-local Misner-Sharp gravitational energy as
\begin{equation}
E_{MS}(x^i):=\frac{1}{2G_N}{\mathcal R}(x^i)\left[1-\chi(x^i) \right]\,.
\end{equation}
For example, the mass of a black hole described by this formalism results to be $E={\mathcal R}_H/(2 G_N)$.
The Killing vector fields $\xi_{\mu}(x^\nu)$ are the generators of the isometries with
$\nabla_{\mu}\xi^{\nu}(x^\nu)+\nabla^{\nu}\xi_{\mu}(x^\nu)=0$: 
in the static case, with the time-like Killing vector field  $K^\mu=\left(1,0,0,0\right)$, the Killing surface gravity $\kappa_K$ is given by
\begin{equation}
\kappa_K K^{\mu}(x^\nu)=K^{\nu}\nabla_\nu K^\mu(x^\nu)\,.
\end{equation} 
In the dynamical case, the real geometric object which generalizes the Killing vector field is 
the Kodama vector field \cite{Kod},
\begin{equation} 
\mathcal K^i(x^i):=\frac{1}{ \sqrt{-\gamma}}\,\varepsilon^{ij}\partial_j{\mathcal R}(x^i)\,,\,i=0,1\,;
\qquad \mathcal K^i:=0\,,\,i\neq 0,1\,.
\end{equation} 
Thus, the Hayward surface gravity associated with dynamical
horizon is \cite{Hay}
\begin{equation}
\kappa_H:=\frac{1}{2}\Box_{\gamma} {\mathcal R(x^i)}\Big\vert_H\,. 
\end{equation}  
The Hawking radiation is a thermal radiation of the black holes due to quantum effects.
In the static case, all derivations of Hawking radiation 
lead to a 
semi-classical expression for the radiation rate $\Gamma$ in terms of the exchange $\Delta E_K$ of the Killing energy $E_K$ and the Killing/Hawking temperature $T_K$,
\begin{equation}
\Gamma\equiv\mathrm{e}^{-\frac{2\pi\Delta E_K}{\kappa_K}}\,,\quad
T_K:=\frac{\kappa_K}{2\pi}\,.
\label{ratekill}
\end{equation}
In the dynamical case 
one may suggest the Kodama/Hayward temperature: 
\begin{equation}
T_H:=\frac{\kappa_H}{2\pi}\,.
\end{equation}
An important example that demonstrates the covariance of the formalism is given by the de Sitter space-time. The static patch reads
\begin{equation}
ds^2=-dt^2(1-H_0^2 r^2)+\frac{d r^2}{(1-H_0^2 r^2)}+r^2 d\Omega^2\,,
\end{equation}
where $\mathcal R=r$ and the horizon is located at $r_H=1/H_0$ with surface gravity $\kappa_H=H_0$. 
The second patch is given by the expanding coordinates of the flat FLRW metric,
\begin{equation}
ds^2=-d t^2+\text{e}^{2H_0 t}\left(dr^2+r^2d\Omega^2\right)\,,\label{flat}
\end{equation}
where $\mathcal R=\text{e}^{H_0 t} r$ and the dynamical (cosmological) horizon is $r_H=1/H_0$ with $\kappa_H=H_0$. Finally, the global patch in non-flat FLRW metric is given by
\begin{equation}
ds^2=-d t^2+\cosh^2 (H_0 t)\left(\frac{dr^2}{(1-H_0^2 r^2)}+r^2d\Omega^2\right)\,,
\label{global}
\end{equation}
with $\mathcal R=r\cosh(H_0 t)$, and $r_H=1/H_0$ and $\kappa_H=H_0$ again.
Now we will see how it is possible to associate a temperature to the dynamical horizon of flat and non-flat de Sitter space-time in (\ref{flat})--(\ref{global}).

\section { Quantization of massless field in FLRW metric}

We recall the quantization of a conformal coupled massless scalar field in the FRLW space-time. The metric reads
\begin{equation}
ds^2=a^2(\eta)(-d\eta^2+d \Sigma_3^2)\,,\quad d\Sigma_3^2= \frac{dr^2}{1-kh_0^2 r^2}+r^2dS_2^2\,.\label{metric}
\end{equation}
where $d\eta=d t/a(t)$ is the conformal time, $ h_0$ is a mass scale and the topology of the spacial section can be
flat, spherically or hyperbolic for $k=0,1,-1$, respectively.

Given a massless scalar  field,
\begin {equation}
\phi(x)=\sum_{\alpha}f_{\alpha}(x)a_{\alpha}+f^*_{\alpha}(x)a^+_{\alpha}\,,
\end {equation}
such that the modes are conformal invariant, namely $(\Box-R/6) f_\alpha(x)=0$,
the associated Wightman function $W(x, x')=<\phi(x)\phi(x')>$ results to be
\begin{equation}
W(x,x')=\sum_{\alpha} f_{\alpha}(x)f^*_{\alpha}(x')\,,\quad\left(\Box -\frac{\mathcal R}{6}\right)W(x, x')=0\,. 
\end{equation}
The Wightman function satisfies the following rule for the conformal transformations of the metric:
\begin{equation}
ds^2=\Omega(x)^2 ds_0^2\,, \quad \phi=\frac{1}{\Omega}\phi_0\,,\quad W(x,x')=\frac{1}{\Omega(x)\Omega(x')}W_0(x,x')\,.\label{conf}
\end{equation}
We may also take $W(x,x')=W(\eta-\eta', r-r')$ due to the homogeneity and isotropy of FLRW space-times. 

Let us consider the spherical case ($k=1$) in (\ref{metric}),
\begin{equation}
ds^2=a^2(\eta)\left( -d\eta^2+d\chi^2+\frac{1}{h_0^2}\sin^2 h_0\chi dS^2_2 \right)\,, 
\quad 
h_0\chi=\arcsin h_0 r\,.
\end{equation}
This  metric  is conformally related to the Minkowski space-time,
\begin{equation}
ds^2=a^2(\eta)4\cos^2 \left(h_0\frac{\eta+\chi}{2}\right)4\cos^2 \left(h_0\frac{\eta-\chi}{2}\right)\left( -dt^2+dr^2+r^2 dS^2_2 \right)\,,
\end{equation}
with
\begin{equation}
t \pm r=\frac{1}{h_0}\tan \left(h_0\frac{\eta \pm \chi}{2}\right)\,.
\end{equation}
Thus, by starting from the well-known Wightman function in Minkowski space-time, one can use (\ref{conf})  and derive for the spherical FLRW metric
\begin{equation}
W(x, x')=\frac{h_0^2}{8 \pi^2 a(\eta) a(\eta')}\,\frac{1}{\cos(h_0(\eta-\eta'))-\cos(h_0(\chi-\chi'))}\,.\label{Wsph}
\end{equation}
The hyperbolic case $k=-1$ is obtained with the substitution $h_0 \rightarrow i h_0$, while the flat case $k=0$ corresponds to the limit $h_0 \rightarrow 0$. 

\section{Quantum fluctuations in flat space-time}

Let us consider a free massless quantum scalar field $\phi(x)$ in thermal equilibrium at the temperature $T$ 
in flat space-time. We know that finite temperature field theory effects of this kind can be investigated by given that the scalar field defined in the Euclidean manifold $S_1 \times R^3$, where the imaginary time is $\tau=-it$, compactified in the circle $S_1$ with period $\beta=1/T$.

We briefly review the local quantity $<\phi(x)^2>$, which is a divergent quantity due to the
product of valued operator distribution in the same point $x$. By making use of the zeta-function regularization procedure, the quantum fluctuations read \cite{haw76,byt96}:
\begin {equation}
<\phi(x)^2>=\zeta(1|L_\beta)(x)\,, \quad L_\beta=-\partial^2_\tau-\nabla^2\,,
\end {equation}
where $ \zeta(z|L_\beta)(x)$ is the local zeta-function associated with the operator $L_\beta$. 
 It is easy to see that the analytic continuation of the
local zeta-function is regular at $z = 1$ and finally one gets
\begin{equation}
<\phi(x)^2>=\frac{1}{12 \beta^2}=\frac{T^2}{12}\,.
\end{equation}
In this way we obtain the temperature of the quantum field in thermal equilibrium from
the zeta-function renormalized  vacuum expectation value, namely we have a quantum thermometer.

\subsection{Quantum fluctuations in FLRW space-time}

Now we extend the argument to generic FLRW metric.
The off-diagonal Wightman function (\ref{Wsph}) leads to
\begin{equation}
W(x,x')=<\phi(x)\phi(x')>=\frac{1}{4\pi^2}\frac{1}{\Sigma^2(x,x')}\,,
\end{equation}
with
\begin{equation}
\Sigma^2(\tau,\tau-s)= a(\tau)a(\tau-s)\frac{2}{h^2_0}\left(\cos h_0(\Delta \chi(s))-\cos h_0 (\Delta \eta(s))\right) \,, 
\end{equation}
where $a(\tau)$ is the conformal factor. Thus, in the limit $s \rightarrow 0$, one has
\begin{equation}
<\phi(x)^2>=W(\tau, \tau)\,.
\end{equation}
It is possible to show that 
\begin{equation}
W(\tau,\tau-s)=-\frac{1}{4\pi s^2}+\frac{B}{48\pi^2}+O(s^2)\,,
\end{equation}
where
\begin{equation}
B=H^2+A^2+2\dot H\dot t+\frac{h_0^2}{a^2}(1-2\dot t^2)\,,\quad
A^2=\frac{1}{\dot t^2-1}\left(\ddot t+H(\dot t^2-1)\right)\,,
\end{equation}
the dot being the derivative respect to the proper time, $H=(d a(t)/dt)/a(t)$ being the usual Hubble parameter, and $A^2$ the radial acceleration. Therefore, after the regularization for the divergent part 
at $s\rightarrow 0$, 
\begin{equation}
<\phi^2>|_R=\frac{1}{48\pi^2}\left(H^2+A^2+2\dot H \dot t\pm\frac{h_0^2}{a^2}(1-2\dot t^2)\right)\,.
\end{equation}
This result is quite general and it is valid also for spatial curvature $k\neq 0$.

\subsection{Quantum fluctuations in non flat de Sitter space-time}

In the case of de Sitter space-time with $k=1$, we may put $H_0=h_0$ and the expression for  quantum fluctuations reads
\begin{equation}
<\phi^2>_R=\frac{1}{48}\left(H_0^2+A^2\right)=\frac{1}{48\pi^2}\frac{H_0^2}{1-R_0^2 H_0^2}\,,
\end{equation}
where $R_0=\text{const}$ is the areal radius of the Kodama observer and the acceleration has been computed as
\begin{equation} 
A^2=\frac{R_0^2H_0^4}{1-R_0^2 H_0^2}\,.
\end{equation}
For a Kodama observer with $R_0=0$ we recover the Gibbons-Hawking temperature associated with de Sitter space-time,
\begin{equation}
T=\frac{H_0}{2\pi}\,.
\end{equation}
This is an important check of our approach, since it shows the coordinate independence of the result for the important case of de Sitter space-time.

\subsection{Quantum fluctuations in FRLW form of Minkowski space-time}

The Minkowski space-time may be written in a FRLW form with hyperbolic section $k=-1$ (Milne universe),
\begin{equation}
ds^2_M=-dt^2+t^2\left(\frac{dr^2}{1+r^2}+r^2 d\Omega^2_2\right)\,,\quad h_0=1\,.
\end{equation}
Making use of the Hayward formalism, it is easy to verify that there is no dynamical horizon and the surface gravity is vanishing. In this case we obtain
\begin{equation}
<\phi^2>_R=\frac{A^2}{48 \pi^2} \,,
\end{equation}
namely only the radial acceleration $A^2$ is present and the temperature is defined as  
\begin{equation}
T_U=\frac{A}{2\pi}\,,
\end{equation}
recovering the well known Unruh effect.

\section*{Acknowledgments}
 This proceeding is based on the paper \cite{Rab}. I thank Sergio Zerbini for helpful discussions.

\end{document}